\documentclass[namedreferences]{myapss}
\usepackage{epsfig}

\begin{document}

\begin{article}

\begin{opening}

\title{THE NEARBY FIELD GALAXY SURVEY}
\subtitle{A spectrophotometric and photometric study of nearby galaxies}

\author{ROLF A. \surname{JANSEN}\protect{$^{1,2,}\footnotemark[1]$}
   \email{rjansen@astro.estec.esa.nl}
   \footnotetext[1]{presently at the Astrophysics Division, Space
   Science Dept.\ of ESA, ESTEC, Postbus 299, NL-2200 AG Noordwijk, The
   Netherlands}}
\institute{\protect{$^1$} Kapteyn Astronomical Institute,
   Postbus 800, NL-9700 AV Groningen, The Netherlands}
\author{SHEILA J. \surname{KANNAPPAN}\protect{$^2$}
   \email{skannappan@cfa.harvard.edu}}
\institute{\protect{$^2$} Harvard-Smithsonian Center for
   Astrophysics, Cambridge, MA 02138, U.S.A.}

\begin{abstract}

\noindent {\bf Abstract.} We report on our observing programme to obtain
integrated spectrophotometry, intermediate and high-resolution
major-axis spectra, and $UBR$ surface photometry of a representative
sample of $\sim$200 galaxies in the nearby field.  The main goal of this
programme is to provide a comparison sample for high-redshift studies
and to study the variation in star formation rates (SFRs), star
formation history (SFH), excitation, metallicity, and internal
kinematics over a wide range of galaxy luminosities and morphological
types.  In particular, we extend the work of Kennicutt (1992) to
lower-luminosity systems. 

We present the main results of our analysis sofar.  In these
proceedings, we condense the presented two atlases of 1) images and
radial surface brightness profiles and colour profiles, and 2) images
and integrated spectra into several example images, profiles and
spectra, showing the general trends observed.  For the original atlasses
we refer the reader to the electronic distribution on CDROM or as
available on the Web at http://www.astro.rug.nl/$\sim$nfgs/\/. 

\end{abstract}

\end{opening}

\section{Introduction}

As galaxies are now routinely sampled at fainter magnitudes and higher
redshifts than ever before, one of the major problems with the
interpretation of distant spectroscopic data has become the difficulty
of obtaining good comparison samples in the local Universe.  Distant
galaxies subtend a small angle on the sky and their spectra are
unavoidably integrated spectra, while most spectra of nearby galaxies
are nuclear spectra only.  A direct comparison of distant and nearby
galaxy spectra, therefore, is difficult.

In a pioneering effort, Kennicutt (1992) obtained integrated
spectrophotometry for 90 galaxies spanning the entire Hubble sequence. 
His study has been a benchmark for the interpretation of spectra at both
high and low redshift.  The range in luminosity sampled per type,
however, was limited to the brightest galaxies, and no uniform surface
photometry or internal kinematic data is available.  Also, only 44 out
of 90 galaxies were observed at intermediate spectral resolution
(5--7\AA), the remainder at lower resolution (15--20\AA).  Large
homogeneous samples of intermediate- or high-resolution nuclear and
integrated spectrophotometry, supplemented by multifilter surface
photometry, for galaxies spanning the entire Hubble sequence and with a
large range in luminosity, are absent in the literature to date.

\section{The Nearby Field Galaxy Survey}

With our Nearby Field Galaxy Survey (NFGS) we aim to remedy this
situation.  The purpose of our study is to obtain integrated and nuclear
spectrophotometry over the entire optical regime, as well as $U$, $B$,
$R$, $H$ and $K'$ surface photometry, and high-resolution spectroscopy,
for a sample of 196 galaxies in the nearby field, including galaxies of
all types and spanning a large range in luminosity.  By `field' we imply
a selection that includes galaxies in clusters, groups an low-density
environments, as opposed to a selection favoring any single one of
these. 

The data will be used to study the emission- and absorption-line
strengths, star formation rate and star formation history, morphologies,
structural parameters, colours, magnitudes and internal kinematics of
gas and stars, both globally and as a function of radius within a
galaxy.  We thus aim to extend the work of Kennicutt to lower-luminosity
galaxies accross the Hubble sequence and to study the variation in
galaxy properties over a wide range of absolute magnitudes and types. 

The data can be used as a benchmark for galaxy evolution modelling and
comparison with observations of high-redshift galaxies, as will result
from future observations with large ground-based telescopes and the
NGST.

\section{Selection of the sample}

The 196 target galaxies in this survey have been objectively selected
from the CfA redshift catalog (CfA~I, Huchra {\it et al.}, 1983)
to span the full range in absolute $B$ magnitude present in CfA~I
($-14<M_B<-22$), while sampling fairly the changing mix of morphological
types as a function of luminosity.  Absolute magnitudes were calculated
directly from blue photografic magnitudes and radial velocities in the
Local Group restframe, assuming $H_0 = 100$ km s$^{-1}$ Mpc$^{-1}$. 

To avoid a sampling bias favouring a cluster population, we excluded
galaxies in the direction of the Virgo Cluster with velocities smaller
than 2000 km sec$^{-1}$.  We also minimized the number of galaxies
larger than 3 arcmin (the slit length of the FAST spectrograph) by
imposing a luminosity-dependent lower limit on the radial velocity. 
Thus, we do not impose a strict diameter limit, while avoiding selecting
the nearest high-luminosity galaxies (which tend to be the largest on
the sky).  We sorted the 1006 galaxies remaining after our Virgo Cluster
and radial velocity cuts into 1-magnitude-wide bins of absolute
magnitude, which in turn were sorted according to Hubble type.  We then
drew from each bin a number of galaxies chosen to approximately
reproduce the local galaxy luminosity function, while preserving the mix
of morphological types in each luminosity bin.  The total number of
galaxies selected is 196, with a median redshift of 0.01 and a maximum
redshift of 0.07\/.  Only eight of these have major-axis optical
diameters larger than the slit length. 

\begin{figure}[t]
\centering
\mbox{
   \epsfig{file=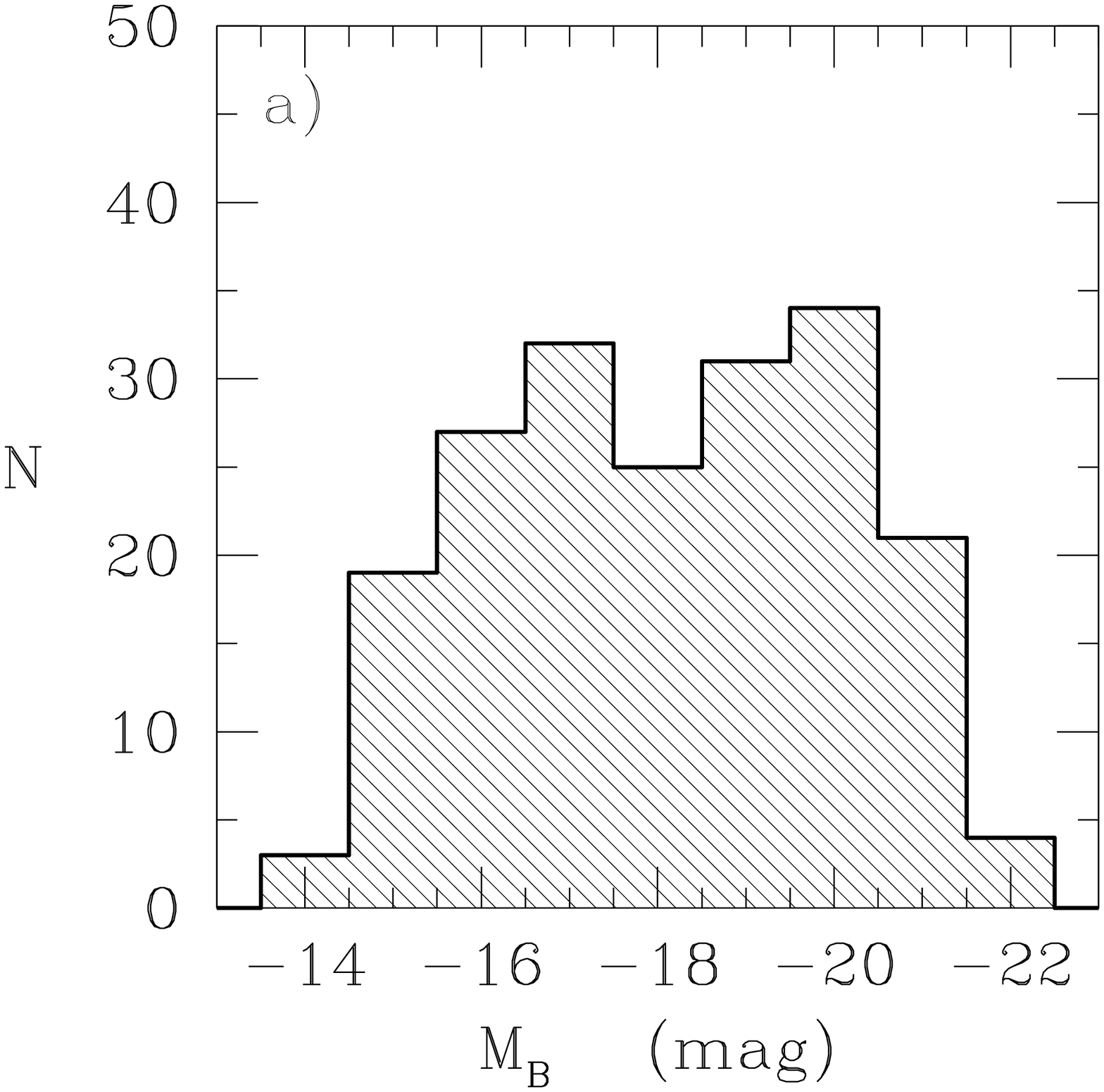,width=0.45\textwidth}
   \epsfig{file=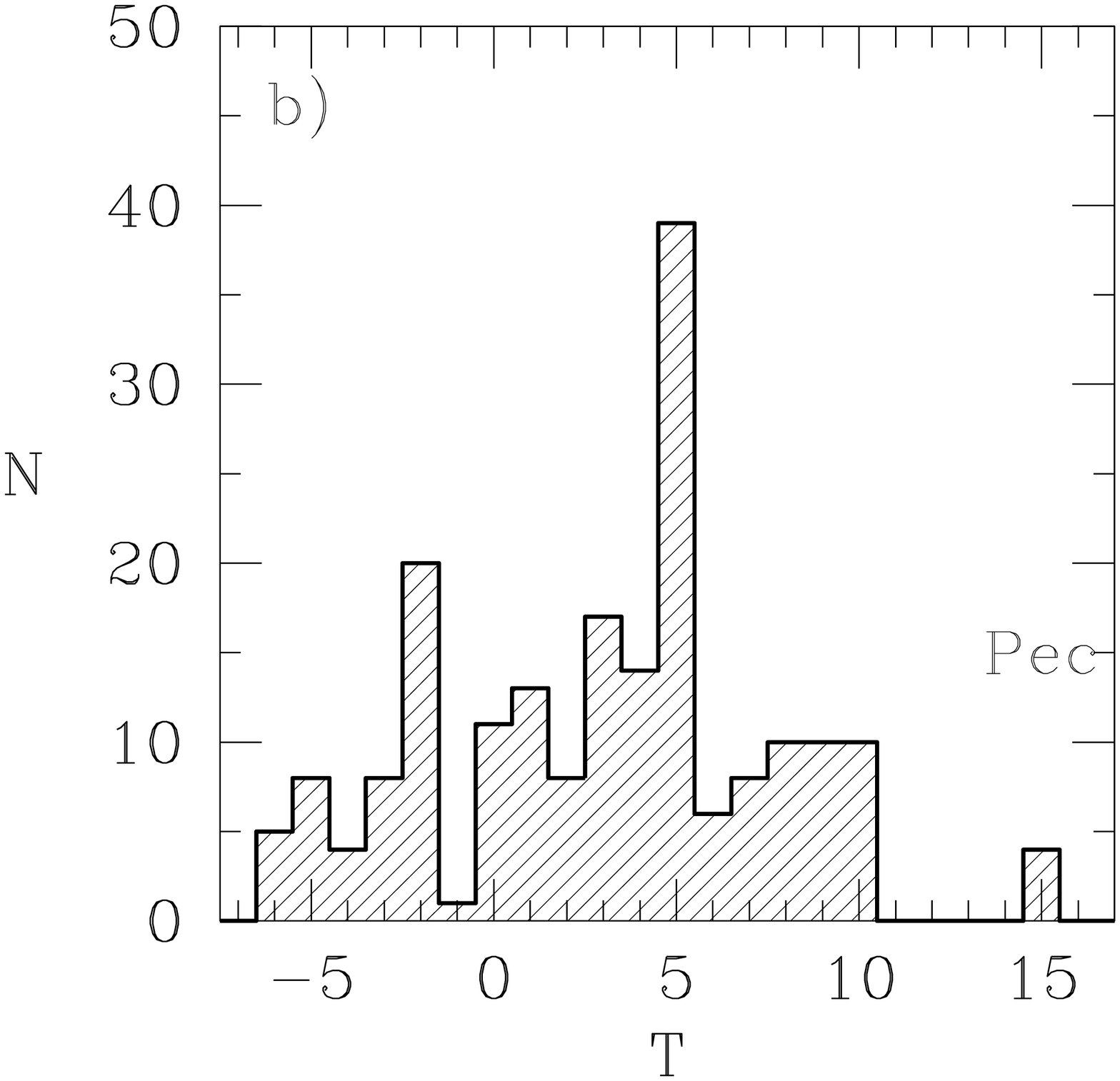,width=0.45\textwidth}
}\par
\mbox{
   \epsfig{file=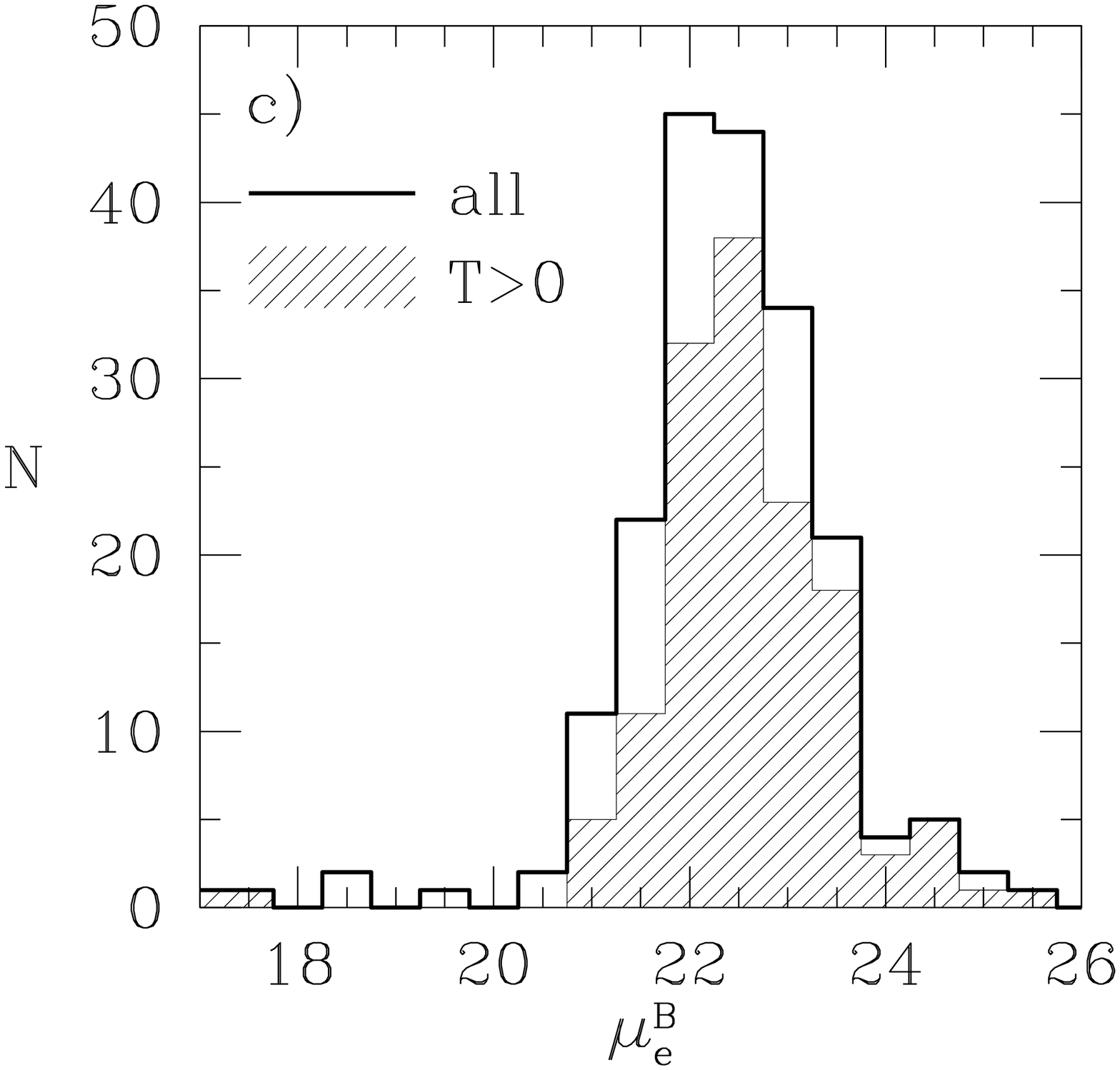,width=0.45\textwidth}
   \epsfig{file=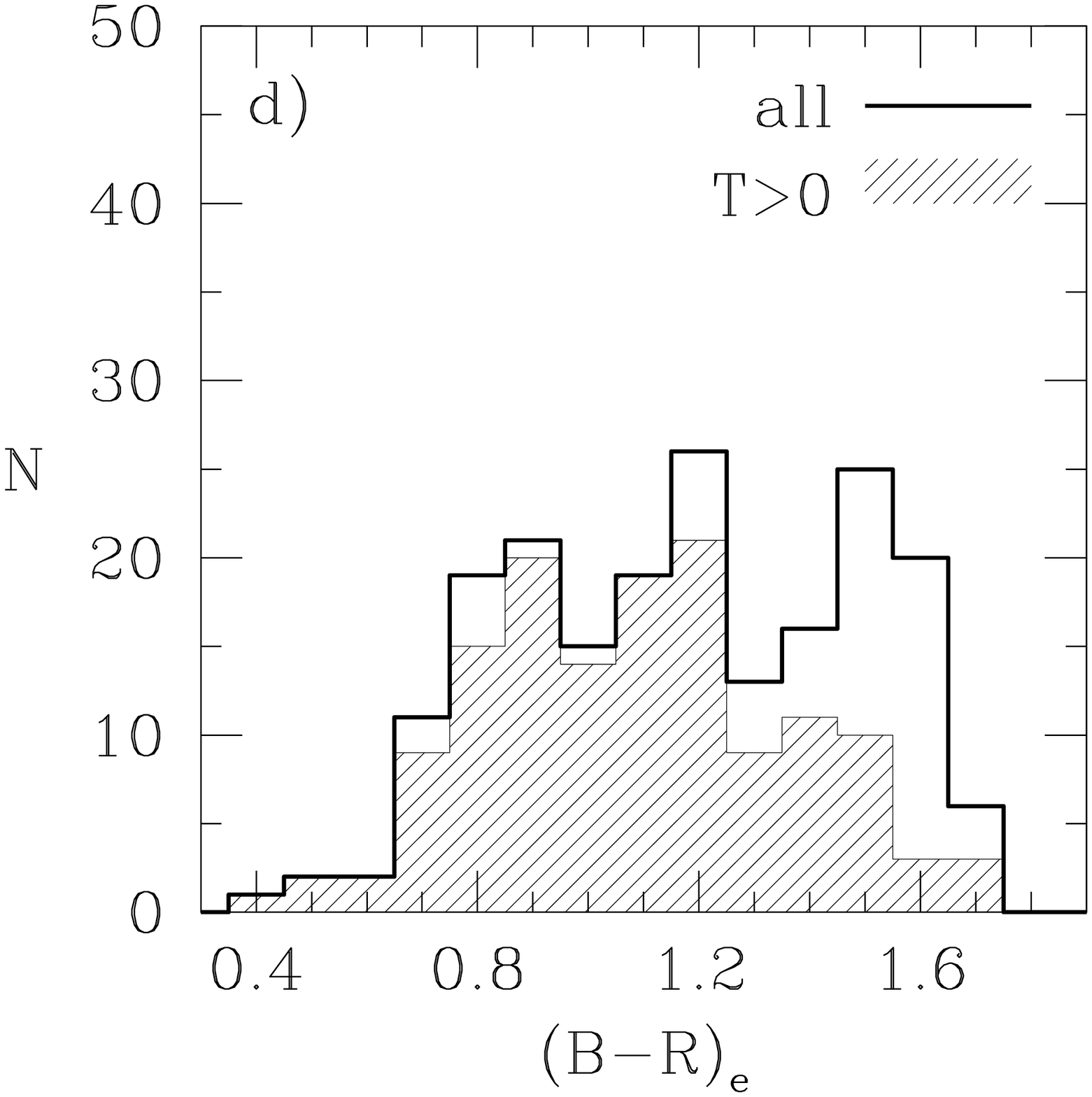,width=0.45\textwidth}
}
\caption{Overview of the global properties of the selected galaxy
sample. Presented are the number distributions as a function of: {\it
a)} absolute $B$ magnitude; {\it b)} numerical galaxy type (note that we
choose to place the unclassified early-type galaxies at $T=-4$ (cD)
rather than $-7$, as we do not have any cD galaxies in our sample, and
this placing seems more natural with respect to the compact ellipticals
at $T=-6$); {\it c)} the surface brightness at the effective radius in
$B$; and {\it d)} the effective $B-R$ colour, measured within that
effective radius. In panels {\it c)} and {\it d)} the open histograms
represent the selected sample, and the shaded ones the subsample of
galaxies with types later than S0/a.}
\end{figure}

In Figure~1 we give an overview of the global properties of the selected
galaxy sample.  The number distributions in magnitude, type and colour
are broad and, although the distribution in effective surface brightness
is peaked, we do have several examples of low surface brightness
galaxies. 

With very few caveats this sample can be considered a fair
representation of the local galaxy population.

\section{Observations}

\subsection{$UBR$ Surface photometry}

The photometric observations were obtained at the F.L.~Whipple
Observatory's 1.2 m Telescope\footnote{The F.L.~Whipple Observatory
(FLWO) is operated by the Smithsonian Astrophysical Observatory and is
located on Mt.~Hopkins in Arizona.} during 50 dark nights between 1994
March and 1997 March.  $U$-filter images were obtained with a thinned
back-illuminated CCD, while most $B$ and $R$ images were obtained with
an older camera and a front-illuminated CCD.  Typically, we exposed
$2\times450$, $2\times900$ and $1\times 900$ seconds to reach limiting
surface brightnesses of $\mu_R=26.1$, $\mu_B=27.2$, and $\mu_U=26.7$ mag
arcsec$^{-1}$ in $R$, $B$, and $U$. 

Radial surface brightness profiles were extracted using an ellipse
fitting procedure (J\o rgensen, Franx \& Kj\ae rgaard, 1992) with
centre, position angle and ellipticity fixed at all radii to the average
values in the outer parts.  Total, effective and isophotal magnitudes
were calculated from the radial surface brightness profiles.  Tests of
the internal and external accuracy of our photometry indicate typical
errors of 0.02 mag and 0.05 mag for isophotal and total magnitudes,
respectively.

\subsection{Spectrophotometry}

Integrated and nuclear spectra were obtained with the FAST spectrograph
(Fabricant {\it et al.}, 1998) at the FLWO 1.5 m Telescope during 41
nights between 1995 March and 1997 March.  The $2720\times512$ pixel
thinned CCD of the FAST in combination with a 300 l/mm grating allowed
coverage of the entire near-UV through optical range (3500--7250\AA) in
a single exposure, at a resolution (FWHM) of $\sim$6\AA.  We aligned the
spectrograph slit approximately along the major axis of each galaxy. 
The integrated spectra were obtained by drift-scanning the slit over a
total distance of half the minor-axis optical diameter and extracting
over 0.7 times the major-axis optical diameter (Figure~2).  Thus, on
average, we sample almost 80\% of the total galaxy light. 

Based on a comparison of the $B-R$ colours measured in our photometry
and the synthetic colours measured in the spectra, and on a comparison
with several galaxies from Kennicutt's sample reobserved for this goal,
we claim an overall relative spectrophotometric accuracy of 6\% (Jansen
{\it et al.}, 2000b). 

\begin{figure}[t]
\centering
\mbox{
   \epsfig{file=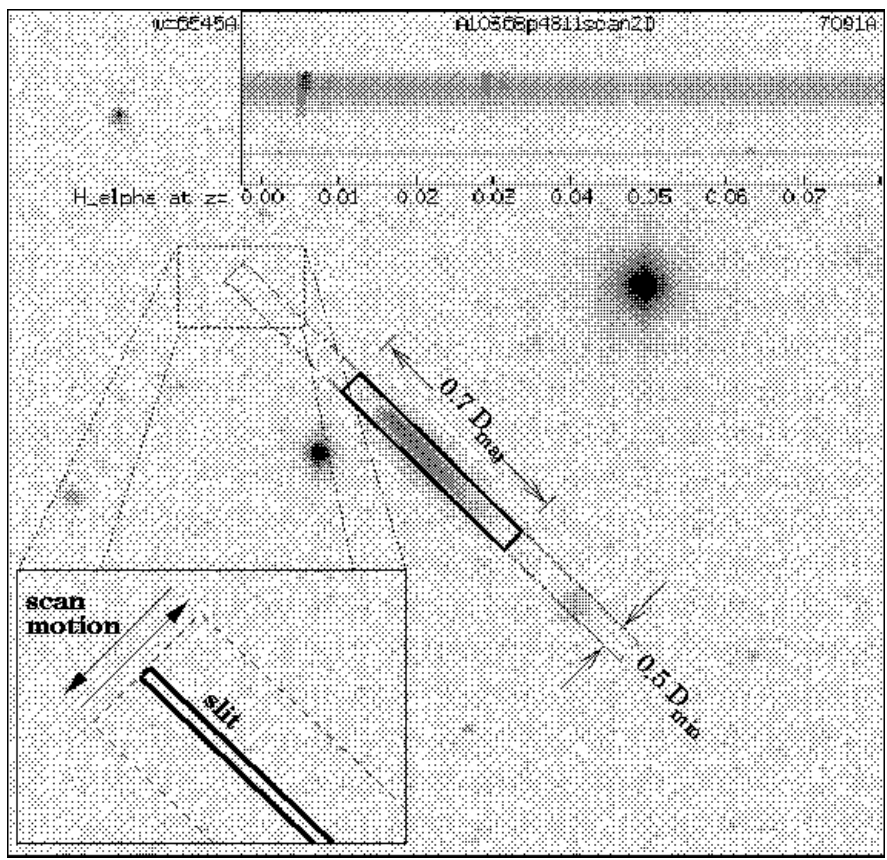,width=0.85\textwidth}
}
\caption{Example of the geometry of the periodic drift scan to obtain
integrated spectra. The total distance over which the slit is moved back
and forth is half the blue minor-axis optical diameter. This distance
was chosen to match the expected surface brightness limit of the 1.5~m
Telescope and FAST spectrograph. The spectra were extracted into 1D
spectra using an objectively defined aperture of size 0.7 times the
major-axis diameter at $\mu_B=26$ mag arcsec$^{-1}$, as determined from
our $B$-filter photometry. The ellipse drawn in the image was fitted to
the $B_{26}$ isophote. In the case of galaxy A10368+4811, we sample
$\sim$80\% of the light within this isophote, or 68\% of the total
galaxian light.
}
\end{figure}

\subsection{Internal kinematics}

Kannappan (2001) has obtained high-resolution spectra in the range
6000--7000\AA\ (emission-line galaxies) at a resolution (FWHM) of 1\AA,
and in the range 4000--6000\AA\ (no or relatively little emission) at a
resolution (FWHM) of 2.3\AA.  Rotation curves were derived from the
former (see Figure~3) by simultaneously fitting the wavelengths of
H$\alpha$, [N~{\sc ii}], [S~{\sc ii}] and [O~{\sc i}] lines as a
function of radius in a galaxy.  Velocity dispersions were fit using a
velocity-broadened stellar template and a Fourier fitting algorithm
(Franx, Illingworth, \& Heckman, 1989) in the latter spectra. 

The main goals of these measurements are to study galaxy mass profiles
as a function of morphological and spectrophotometric properties, to
compare gas and stellar kinematics and investigate kinematic evidence of
galaxy interactions, mergers and mass infall. 

\begin{figure}[t]
\centering
\mbox{\epsfig{file=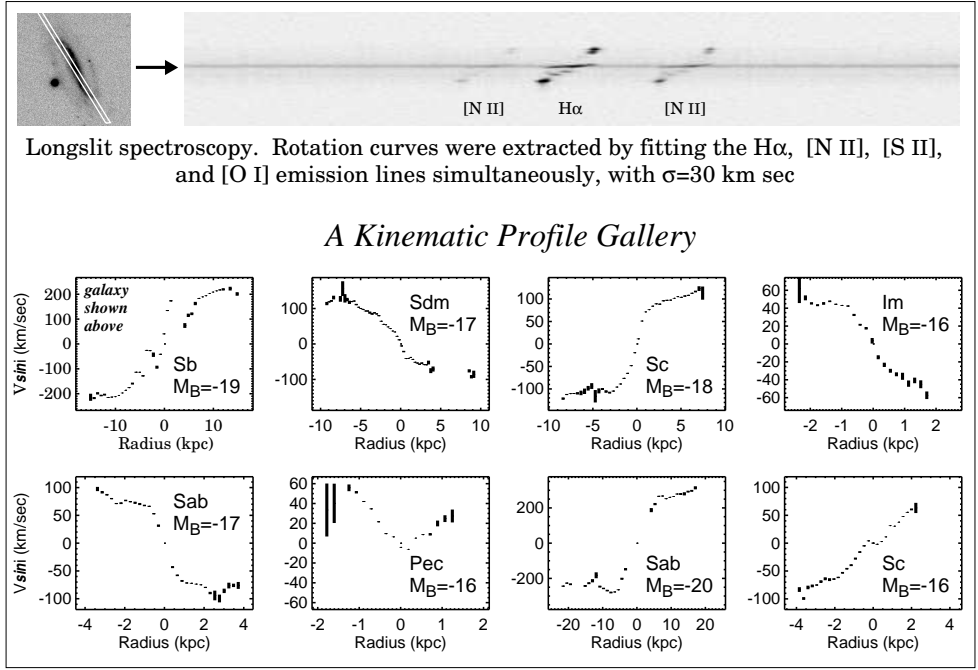,width=0.95\textwidth}}
\caption{Gallery of several example optical rotation curves. At {\it top
left}, a greyscale $B$-filter image is given for one galaxy with the 
orientation of the spectrograph slit indicated. The resulting 2D
spectrum image, spanning the range 6000--7000\AA\ in wavelength, is
presented with the strongest emission lines marked. The extracted
rotation curve for this galaxy is presented in the {\it upper left
panel} of the gallery, showing that this galaxy is similar in size,
luminosity and mass to our own Milky Way galaxy.}
\end{figure}

\subsection{$H$ and $K'$ surface photometry}

Pahre {\it et al.\/} (in prep.) have obtained near-infrared $H$ and $K'$
filter surface photometry for half of the NFGS galaxies.  The main
purpose of these observations is to determine the relative contributions
of spheroids and disks to the luminosity density of the local universe. 

Previous measurement of this parameter (Schechter \& Dressler, 1987)
relied on visual total magnitude and bulge estimates from photographic
plates, and suffered from relatively poor knowledge of the luminosity
functions of different morphological types.  Combination of $U$, $B$,
$R$, $H$ and $K'$ photometry and our spectrophotometry will allow a
better separation of age and metallicity effects, as well as internal
extinction, while inversion of our selection and comparison with a
deeper complete spectroscopic sample ({\it e.g.}, Carter 1999) allows us
to infer volume densities as a function of type and luminosity. 

\newpage

\section{Results}

\subsection{Surface photometry}

We observe a strong trend of $(B-R)_e$ colour with morphological type,
with later-type galaxies becoming progressively bluer (Figure~4{\it a}). 
The observed scatter on this trend is 0.19 mag, which is smaller than
the 0.24 mag scatter on the colour--magnitude trend where intrinsically
fainter systems tend to be bluer than brighter systems (Figure~4{\it
b}).  Estimating a galaxy's broad type class (E, S, Irr) from its colour
can, therefore, be as accurate as estimates based on galaxy asymmetry
and central concentration of the light (Abraham {\it et al.}, 1996). 

\begin{figure}[t]
\centering
\mbox{
   \epsfig{file=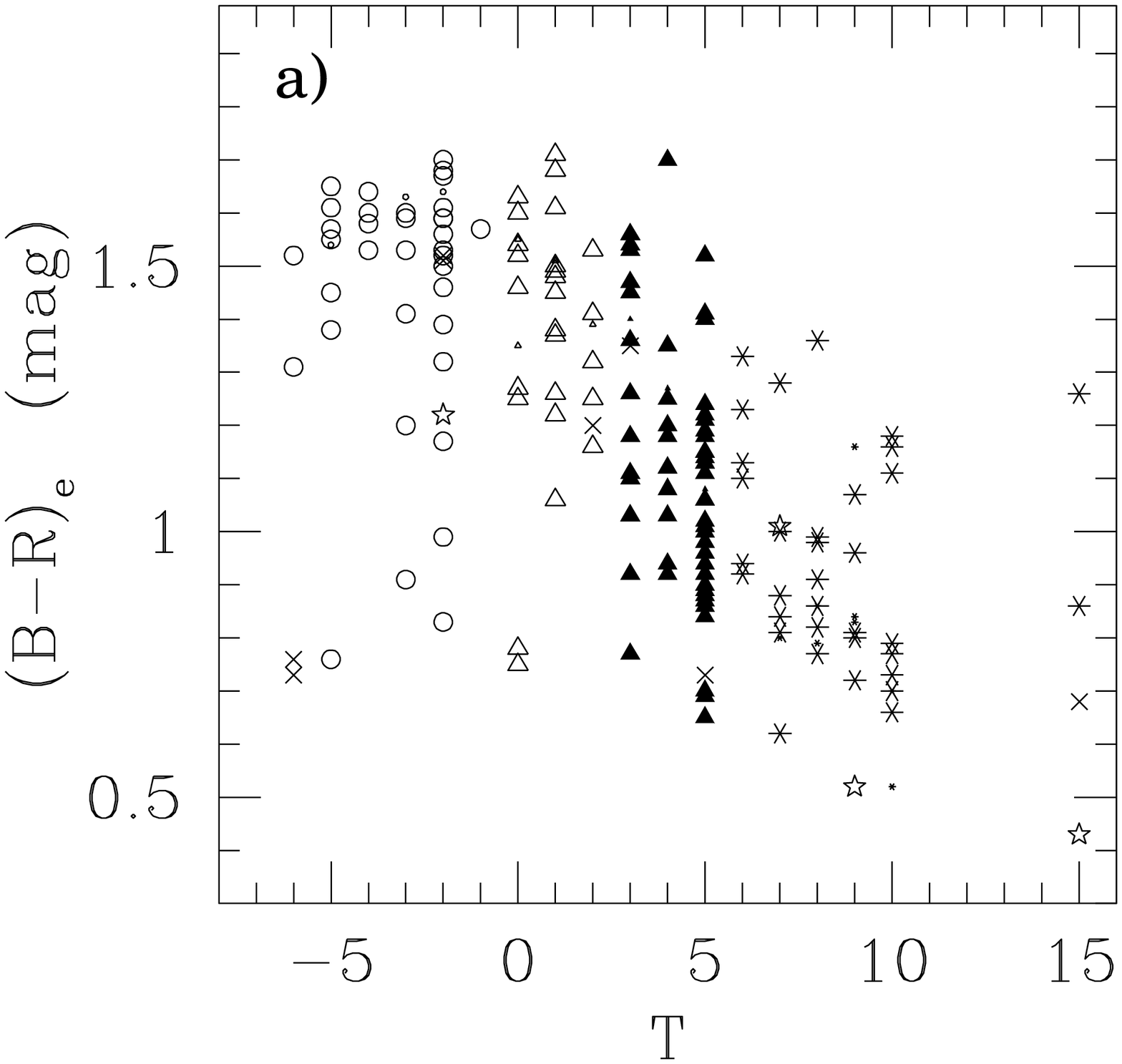,width=0.45\textwidth}
   \epsfig{file=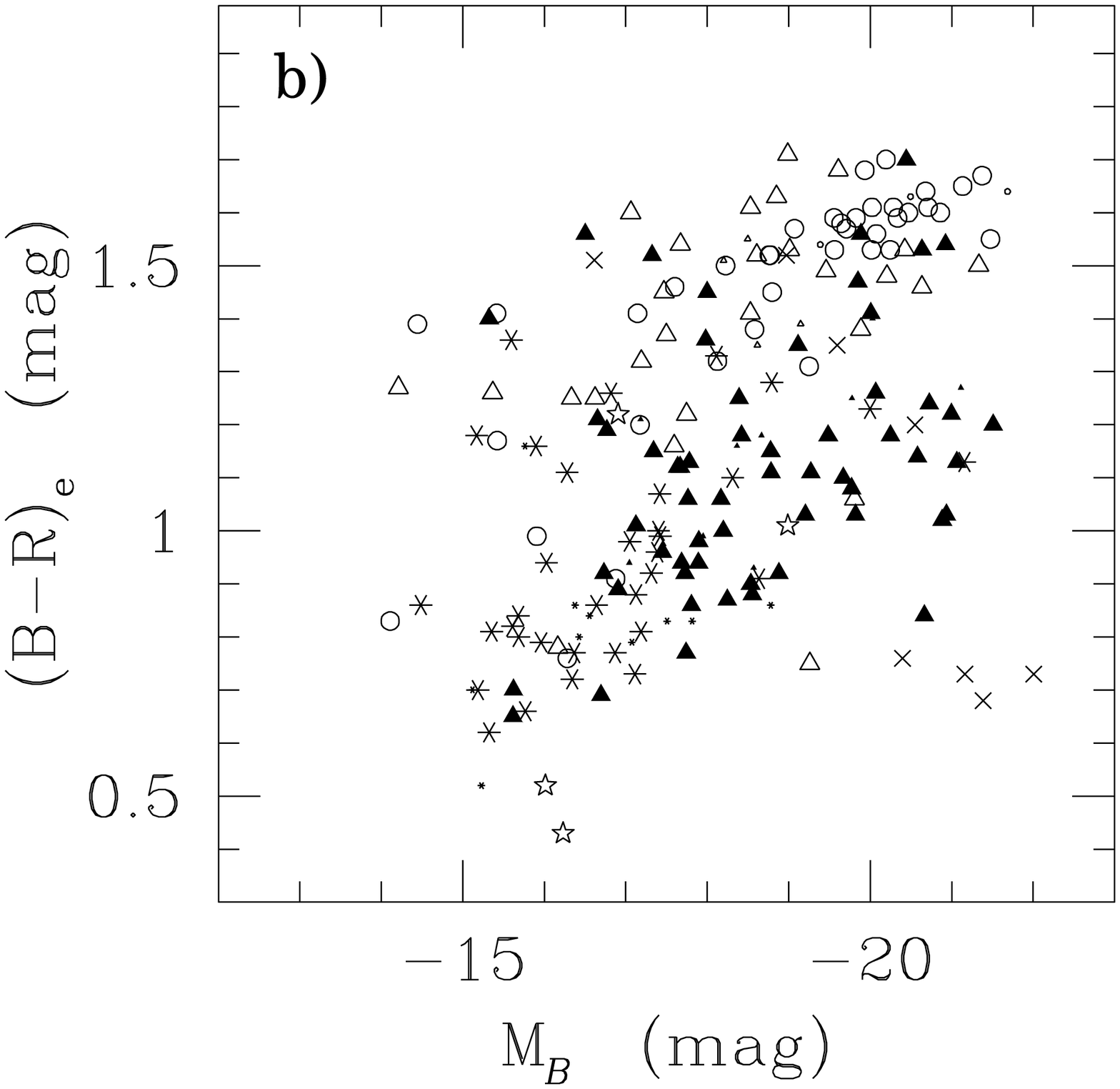,width=0.45\textwidth}
}
\caption{Effective $B-R$ colours as a function of morphological type
and absolute $B$ magnitude.  Points are coded according to type as
follows. Ellipticals and S0s ($T<0$) are indicated by {\it open
circles}, early-type spirals ($0\leq T<3$) by {\it open triangles},
intermediate spirals ($3\leq T<6$) by {\it solid triangles}, and
late-type spirals and irregulars by {\it asterisks}. Active nucleated
galaxies are indicated by {\it crosses} and starburst galaxies by {\it
open stars}.}
\end{figure}

We find that colour--magnitude relations are useful for early-type
systems and to a lesser degree for very late-type systems, but not
useful for intermediate-type spirals.  We also verify the result of
Tully {\it et al.\/} (1996), that the faintest galaxies may become
redder with radius instead of bluer.  Star formation is the driving
force in this trend (Jansen {\it et al.}, 2000a).

\subsection{Spectrophotometry}

The bluing of galaxies towards both later morphological types and lower
luminosities is apparent in the spectrum continua (Figure~5) as well. 
Moreover, the emission line strengths increase with respect to the
continua going to lower-luminosity systems.  This trend is most apparent
for the intermediate-type spiral galaxies, but can be seen all accross
the Hubble sequence. 

\begin{figure}[ht]
\centering
\mbox{\epsfig{file=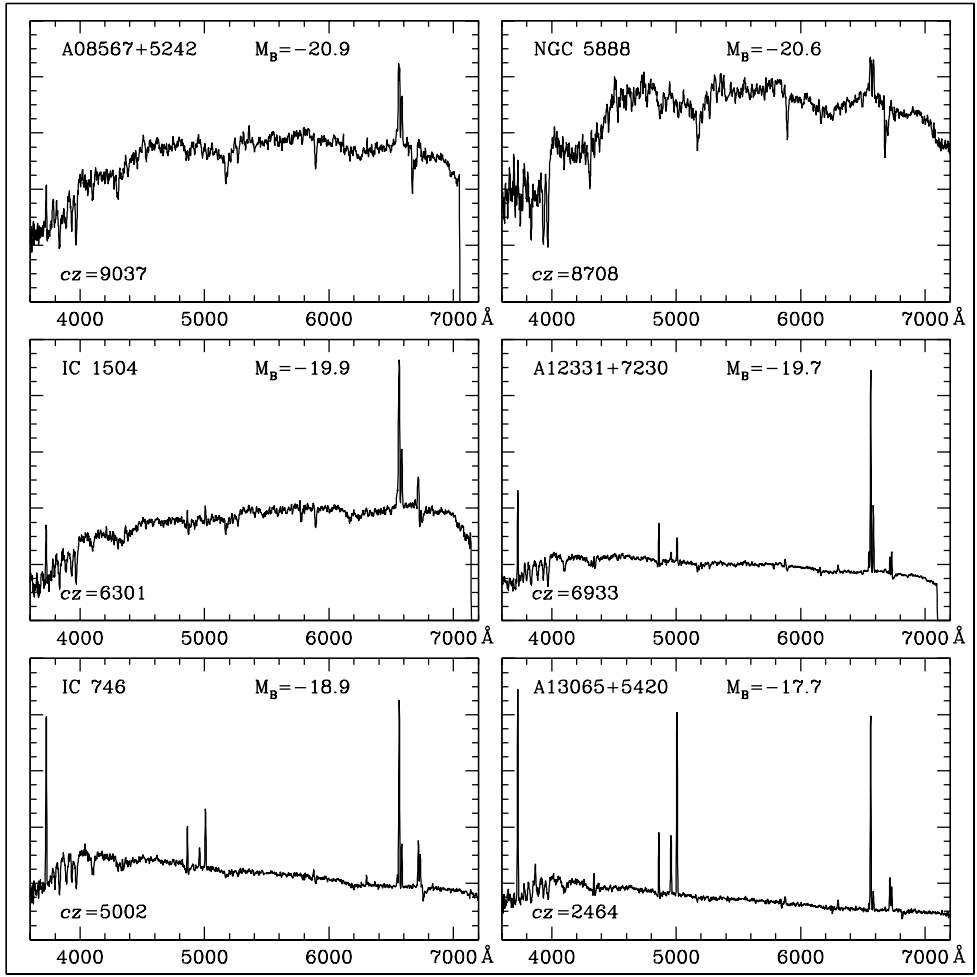,width=0.95\textwidth}}
\caption{Gallery of spectrophotometrically calibrated integrated spectra
as a function of luminosity for six Sb galaxies. For all Hubble types,
continua tend to become bluer and the percentage of galaxies showing
emission increases going to fainter luminosities. The range in spectral
properties at each type and luminosity tends to be large, however.}
\end{figure}

\begin{figure}[ht]
\centering
\mbox{
   \epsfig{file=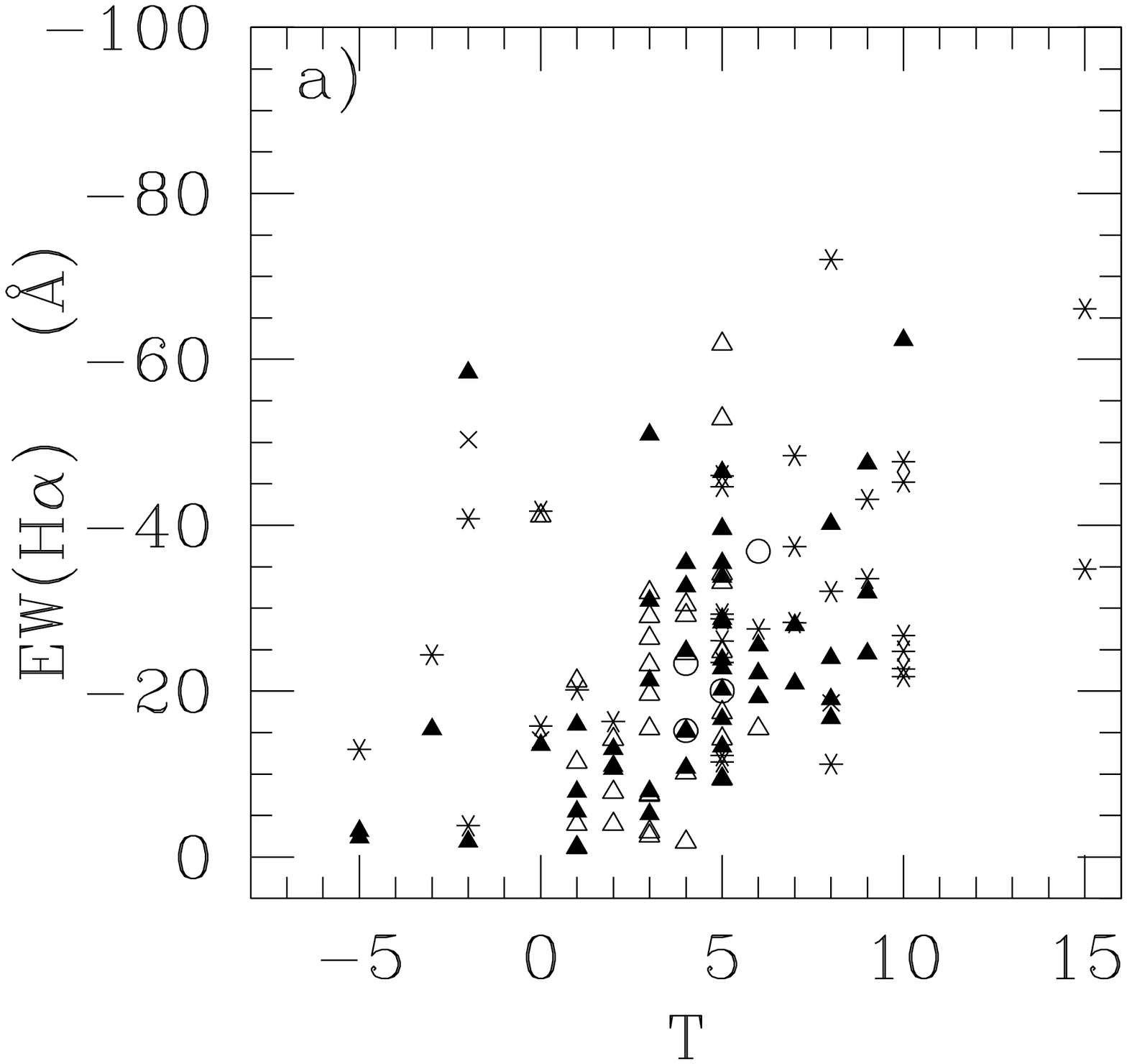,width=0.45\textwidth}
   \epsfig{file=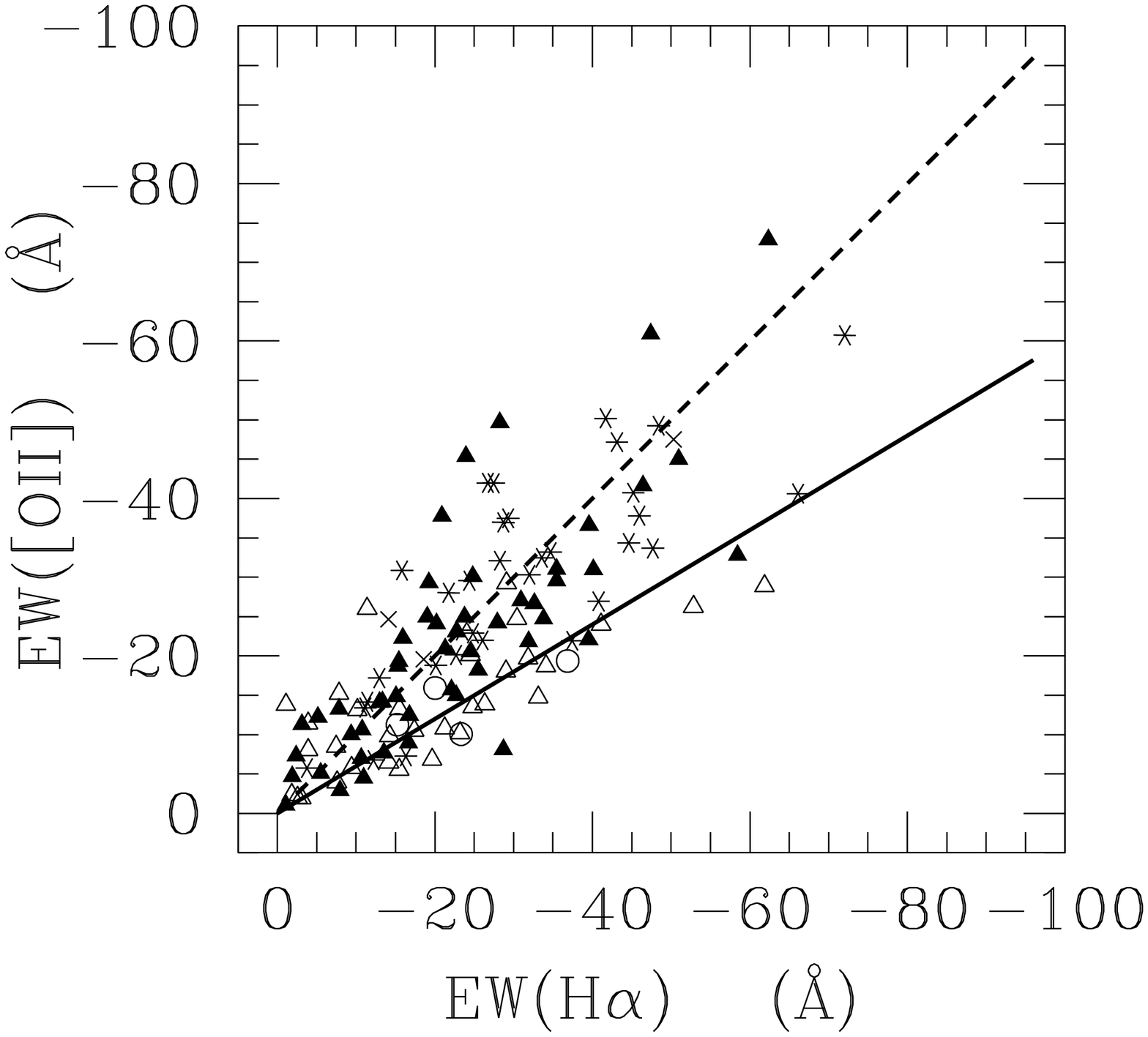,width=0.45\textwidth}
}
\caption{Integrated emission line strengths for the 122 normal star
forming galaxies with EW(H$\alpha$)$>$1\AA. Symbols are coded according
to total $B$-filter luminosity, from high to low, as {\it open circles},
{\it open triangles}, {\it solid triangles}, {\it asterisks} and {\it
crosses}.  {\it a)} H$\alpha$ equivalent widths versus type. {\it b)}
[O~{\sc ii}]3727\AA\ versus H$\alpha$ equivalent widths. High-luminosity
galaxies follow the relation found by Kennicutt (1992; {\it solid
line}), while galaxies fainter than $M_B\sim -19$ tend to follow a much
steeper relation, approximating even EW(H$\alpha$) (see the line of
equality [{\it dashed}]).
}
\end{figure}

Of particular interest is the relative strength of the [O~{\sc
ii}]3727\AA\ line with respect to H$\alpha$, as [O~{\sc ii}] is the
strongest line available in the optical regime for galaxies with
$0.3\lsim z\lsim 1$ and has been widely used as a tracer of the SFR. 
Both figures~5 and 6{\it b} show that the fainter galaxies tend to have
[O~{\sc ii}] in excess of the relation found by Kennicutt (1992),
corresponding to EW([O~{\sc ii}])$\simeq 0.6$\ EW(H$\alpha$).  Jansen
{\it et al.\/} (2001) demonstrated that this must be attributed to a
combination of 1) a smaller amount of interstellar reddening --- which
boosts the strength of [O~{\sc ii}] relative to the Balmer lines over a
large range in temperature --- and 2) a higher average excitation
temperature in the ISM of low-luminosity galaxies than in
high-luminosity galaxies.  The underlying cause for both these trends
was shown to be metallicity.

\end{article}

\end{document}